\documentclass{aa}
\input{psfig}

\catcode`\@=11 
\def\gsim{\ifmmode{\mathrel{\mathpalette\@versim>}}
\else{$\mathrel{\mathpalette\@versim>}$}\fi}
\def\lsim{\ifmmode{\mathrel{\mathpalette\@versim<}}
\else{$\mathrel{\mathpalette\@versim<}$}\fi} 
\def\@versim#1#2{\lower2.9truept \vbox{\baselineskip 0pt \lineskip 0.5truept
\ialign{$\m@th#1\hfil##\hfil$\crcr#2\crcr\sim\crcr}}} 
\catcode`\@=12
\def\lx{$L_{\rm X}\,$}
\def\lup{$L_{\rm X,UP}\,$}
\def\lb{$L_{\rm B}\,$}
\def\mincir{\raise -2.truept\hbox{\rlap{\hbox{$\sim$}}\raise5.truept
\hbox{$<$}\ }}
\def  \magcir{\raise -2.truept\hbox{\rlap{\hbox{$\sim$}}\raise5.truept
\hbox{$>$}\ }} 

\begin{document}
\thesaurus{11.05.1, 11.06.2, 11.09.4, 11.14.1, 13.25.2}

\title{Global X-ray emission and central properties of 
early type galaxies}
\author{S. Pellegrini}
\offprints{S. Pellegrini}
\institute{
Dipartimento di Astronomia, Universit\`a di Bologna,
via Ranzani 1, I-40127 Bologna}
\date{Received...; accepted ...}
\authorrunning{S. Pellegrini }
\titlerunning{Global X-ray emission and central properties of 
E/S0s}
\maketitle
\begin{abstract}

{\it Hubble Space Telescope} observations have revealed that the
central surface brightness profiles of early type galaxies can be
divided into two types: "core" profiles and featureless power law
profiles, without cores.  On the basis of this and previous results,
early type galaxies have been grouped into two families. One consists
of coreless galaxies, which are also rapidly rotating, nearly
isotropic spheroids, and with disky isophotes. The other is made of
core galaxies, which are slowly rotating and boxy-distorted.  Here I
investigate the relationship between global X-ray emission and shape
of the inner surface brightness profile, for a sample of 59 early type
galaxies. I find a clear dicothomy also in the X-ray properties, in
the sense that core galaxies span the whole observed range of \lx
values (roughly two orders of magnitude in \lx), while power law
galaxies are confined to log \lx (erg s$^{-1})<41$.  Moreover, the
relation between \lx and the shape of the central profile seems to be
the strongest among the relations of \lx with the basic properties
characterizing the two families of early type galaxies. As an example,
\lx is more deeply connected with the shape of the central profile
than with the isophotal shape distortion, or the importance of
galactic rotation.  So, a {\it global} property such as \lx, that
measures the hot gas content on a galactic scale, turns out to be
surprisingly well linked to a {\it nuclear} property.

Various possible reasons are explored for the origin of the different
\lx behavior of core and power law galaxies.  While a few explanations
can be imagined for the large spread in the X-ray luminosities of core
galaxies, an open problem is why power law ones never become very
X-ray bright.  It is likely that the presence of a central massive
black hole, and possibly also the environment, play an important role
in determining \lx (i.e., the hot gas content). Therefore the problem
of interpreting the X-ray properties of early type galaxies turns out
to be more complex than thought so far.

\end{abstract}

\section{Introduction}

With {\it Hubble Space Telescope} the central surface brightness
profiles of nearby galaxies can be studied with a resolution limit of
$\sim 0.\hskip -0.1truecm ^{\prime\prime}1$, which corresponds to,
e.g., $\lsim 10$ parsecs in Virgo.  So, a large number of early type
galaxies have been observed with unprecedented detail in their central
regions (e.g., Jaffe et al. 1994, Ferrarese et al. 1994, Lauer et
al. 1995, Forbes et al. 1995; Faber et al. 1997, hereafter F97).  It
turns out that the most luminous galaxies show outer power law
profiles that break internally to shallow inner profiles $I\propto
R^{-\gamma}$, with $\gamma < 0.3$. These galaxies are said to have
cuspy cores\footnote { In the following I use the nomenclature "core
profile" to refer to a profile with a cuspy core, and to "core galaxy"
to refer to a galaxy with a cuspy core.}; their break radii typically
range from 50 to 1000 pc. Faint galaxies show instead steep
featureless power law profiles that lack cores down to the resolution
limit.  Their central surface brightness slope is typically $\gamma >
0.5$, with $\gamma\approx 0.8$ on average.  At intermediate
luminosities core and power law galaxies coexist.

These central properties correlate with isophotal shape and amount of
rotation: core galaxies tend to be boxy and slowly rotating, whereas
power law galaxies tend to be disky and rapidly rotating (F97). At
intermediate luminosities, where power law and core galaxies coexist,
the presence of a core is a better predictor of boxiness or slow
rotation than luminosity; the same can be said for the absence of a
core and diskiness, or high rotation. So, it has been suggested that
early type galaxies can be divided into two families: one includes
those galaxies that are coreless, nearly isotropic spheroids, rotate
rapidly, and have disky isophotes; the other those that have cores,
rotate slowly, are possibly moderately triaxial and boxy-distorted
(Kormendy \& Bender 1996).

In this paper I investigate whether power law and  core early
type galaxies (ellipticals and S0s) differ systematically also in
their soft X-ray luminosities.  I find that power law galaxies are
confined below log $L_X$(erg s$^{-1})=41$, while  core ones
can reach \lx values at least one order of magnitude higher.  This
holds even in the range of optical luminosities where the two families
coexist. So, the central properties of early type galaxies are tightly
related to, and possibly at the origin of, their global X-ray
emission. A key factor could be the presence of a central massive
black hole.  In fact, it has been suggested that nuclear massive black
holes are important for explaining the dicothomy of the inner light
profiles, as they should have substantial influence on the dynamics
and evolution of the surrounding gas and stars (e.g., Cipollina \&
Bertin 1994, van der Marel 1999, Merritt 1999, Nakano \& Makino
1999). I also explore the alternative possibility that the other basic
properties characterizing the two families (importance of rotation and
isophotal shape distortion) affect the global \lx.  This hypothesis
is less plausible because the trend of $L_X$ with these other
properties is less sharp. Finally, I also examine the effects on \lx
that may be produced by differences in the environment.

\section{The sample}

F97 present the fits of the surface brightness
 profiles for a large sample of early type galaxies observed with
 WFPC1; they modeled the data with a double power law with a break
 radius, derived the inner surface brightness slope (the $\gamma $
 parameter) and gave the classification into core or power law galaxy
 accordingly. This classification is given by the same authors for
 nine more early type galaxies in Magorrian et al. (1998).  Crane et
 al. (1993), Ferrarese et al. (1994), Forbes et al. (1995) and Quillen
 et al. (1999) also present {\it HST} central profiles for a number of
 early type galaxies, and fit them with double power laws which govern
 the inner and the outer slopes. For all these galaxies, following
 F97, I have adopted the criterion of classifying them
 as power law galaxies when the central brightness profile does not
 show an inner break, and as core galaxies when it does. In the first
 case the inner slope always turned out to be larger than 0.3, and in
 the second case smaller than 0.3.

I have cross-correlated the sample of Es and S0s with inner profile
measured by {\it HST} with existing catalogs of X-ray emission, namely the
{\it Einstein} based catalog (Fabbiano et al. 1992), the $ROSAT$
all-sky survey based catalog (Beuing et al. 1999), and the list of 61
early type galaxies observed with the $ROSAT$ PSPC by Irwin \& Sarazin
(1998).  So, I have collected a sample of 59 galaxies with information
on both the X-ray emission (detection or upper limit) and the central
surface brightness properties\footnote{The whole sample of galaxies
with information on both the X-ray emission and the central profile
includes three more objects (M87, NGC3862, and NGC6166). These have
been removed because their very high X-ray emission (\lx$>>10^{42}$
erg s$^{-1}$) is known to be largely contaminated by the presence of a
`classical' AGN, while in the present study only normal early type
galaxies are considered (see also Section 5.2). All these three
galaxies have  cores.}. 39 of these objects come from F97, 
9 from Magorrian et al. (1998), 5 from Forbes et
al. (1995), 3 from Quillen et al. (1999), 2 from Ferrarese et al. (1994), 
and 1 from Crane et al. (1993).  The order of priority for taking the 
X-ray information has been as follows: first the Beuing et al. catalog, 
then the Irwin \& Sarazin list, and
finally the {\it Einstein} based catalog.  In the final sample, the
X-ray information comes from the $ROSAT$ all-sky survey catalog for 43
objects, from the Irwin \& Sarazin list for 5 objects, from the {\it
Einstein} based catalog for 9 objects, and for 2 Coma galaxies
(NGC4874 and NGC4889) from the $ROSAT$ study of the Coma cluster by
Dow \& White (1995).

\begin{figure}
\vskip -5truecm
\hskip -1truecm
\parbox{1cm}{
\psfig{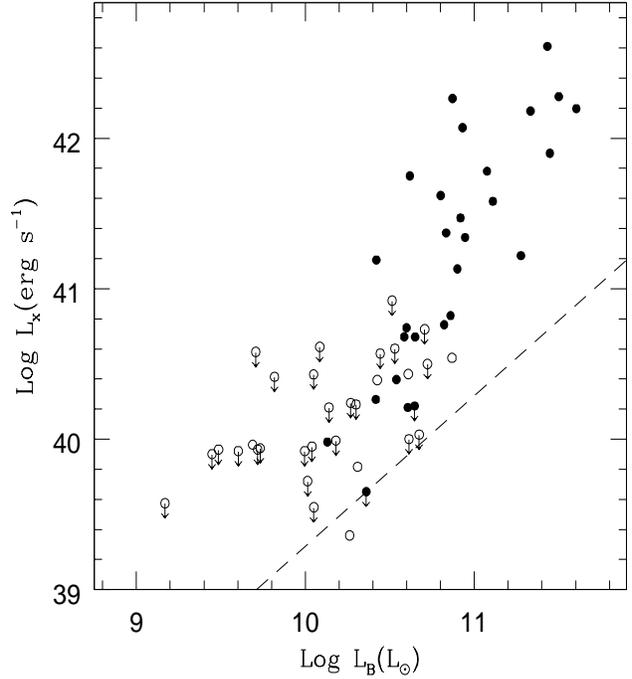} }
\vskip -1truecm
\caption[]{The \lx--\lb diagram for early type galaxies with inner
surface brightness profile measured by $HST$.  Power law galaxies are
shown with open circles and  core galaxies with full
circles. Upper limits on X-ray luminosities are shown with a downward
arrow. The data used are those in Table 1. The dashed line
(\lx$\propto$\lb) is an estimate of the stellar sources contribution
to the X-ray emission (from Kim et al. 1992).}
\end{figure}

\begin{table*}
\caption[] { Galaxy properties}
\begin{flushleft}
\begin{tabular}{ l  r  r  r  r r  l }
\noalign{\smallskip}
\hline
\noalign{\smallskip}
 Name   & d$^a$  & log$L_{\rm X}^b$ & log$L_{\rm B}^c$ &
 100$\cdot a_4/a^d$ & Profile$^e$ & $\gamma^{f} $\\
   & (Mpc)  & (erg s$^{-1})$ & $(L_{\odot})$ & & & \\
\noalign{\smallskip}
\hline
\noalign{\smallskip}
NGC524  &  40.0  & 40.82  &  10.86  &      &  $\cap $       &0.00\\
NGC596 &  39.6 & $<$40.50&  10.72   &  1.30&   $\setminus $ &0.55\\
NGC720  &  35.2  & 41.37  &  10.83  &  0.35&  $\cap $       &0.06 \\
NGC821 &  37.7 & $<$40.73 & 10.66   &  2.50 &  $\setminus $ &\\
NGC1172 &  47.1 & $<$40.92 & 10.51  &      &  $\setminus $  &1.01\\
NGC1052 &  29.3 & 40.68   &  10.59 & 0.00 & $\cap $         & 0.18\\
NGC1316 &  26.4  & 41.22  &  11.28  &  1.00&   $\cap $      &0.00\\   
NGC1399 &  26.8  & 42.26  &  10.87 &  0.10&   $\cap $       &0.07\\
NGC1400 &  31.3  & 40.39  &  10.43 &  0.00&    $\setminus $ &0.35\\
NGC1426 &  30.9 & $<$40.24 & 10.27 &  0.00&  $\setminus $   &0.85 \\
NGC1427 &  26.3  & 39.82  &  10.31 &  0.35 &  $\setminus $  &0.76\\  
NGC1439 &  30.9 & $<$40.23 & 10.30 &  -0.10&  $\setminus $  &0.83\\
NGC1553 &  18.2 & 40.43   &  10.61 &       &  $\setminus $  &0.74\\
NGC1600 &  98.6  & 41.90  &  11.45 & -0.75&  $\cap $        &0.08\\
NGC2300 &  45.6  & 41.62  &  10.80 &  0.55&   $\cap $       &\\
NGC2778 &  37.2 & $<$40.43 & 10.05 & -0.15  &  $\setminus $ &\\
NGC2832 & 136.2  & 42.61  &  11.43 & -0.30&  $\cap $        &0.02\\
NGC3115 &   9.6  & 39.36  &  10.26 &      &  $\setminus $   &0.78\\
NGC3377 &  13.3&  $<$39.72 & 10.01 &  1.05&  $\setminus $   &0.29\\
NGC3379 &  13.2 & $<$39.65 &  10.36 & 0.10&  $\cap $        &0.18\\
NGC3384 &  12.0 & $<$39.55 & 10.05 &      &  $\setminus $   &\\
NGC3599 &  20.4&  $<$39.94 &  9.73 &      & $\setminus $    &0.79\\
NGC3608 &  32.5  & 40.40  &  10.54 & -0.20&  $\cap $        &0.00\\
NGC4168 &  43.8  & 40.68  &  10.65 &  0.37&  $\cap $        &0.14\\
NGC4239 &  21.0 & $<$39.93 &  9.49 &       & $\setminus $   &0.65\\
NGC4278 &  14.6 &  39.98  &  10.13 &  -1.00 &  $\cap $      &\\
NGC4291 &  36.7 &  41.19  &  10.42 &  -0.40 &   $\cap $     &\\
NGC4342 &  27.0 & $<$40.58 &  9.71 &    &  $\setminus $     &1.47\\ 
NGC4365 &  20.2 &  40.21  & 10.61  &  -0.95 &   $\cap $     &0.15\\ 
NGC4374 &  20.7 &  40.76  & 10.82  &  -0.40 &  $\cap $      &0.31\\
NGC4387 &  24.5 & $<$40.41 & 9.82 & -0.75&  $\setminus $    &0.72\\
NGC4406 &  20.8 &  42.07  &  10.93 &  -0.70&  $\cap $       &0.08\\
NGC4434 &  20.3 & $<$39.92 &  9.60 &  0.44 &  $\setminus $  &0.70\\ 
NGC4458 &  20.8  & 39.96  &   9.69 &  0.34 &  $\setminus $  &0.49\\
NGC4464 &  20.3 & $<$39.90 &  9.45 &       & $\setminus $   &0.88\\
NGC4467 &  24.5&  $<$39.57 &   9.17 &      &  $\setminus $  &0.98\\
NGC4472 &  20.3 &  41.78 &  11.08  &  -0.25&  $\cap $       &0.04\\
NGC4473 &  20.8 &  40.26 & 10.42   &  0.90 &  $\cap $       &\\
NGC4478 &  24.5 & $<$40.61 & 10.09 &  -0.75&  $\setminus $  &0.43\\ 
NGC4494 &  22.5 & $<$40.03 & 10.67 &  0.30 &  $\setminus $ & \\ 
NGC4551 &  20.8 & $<$39.93 &  9.72 &  -0.65&  $\setminus $ & 0.80\\
NGC4552 &  20.8 &  40.74  &  10.60 &  0.01 &  $\cap $      &0.00\\
NGC4564 &  20.7 & $<$39.95 & 10.04 &  1.00 &  $\setminus $ & 0.05\\
NGC4589 &  36.6 & $<$40.22 & 10.65 &  0.55 &  $\cap $      &\\
NGC4621 &  20.7 & $<$40.00 & 10.61 &  1.50 &  $\setminus $ &0.50\\
NGC4636 &  19.9 &  41.75  & 10.62  & -0.10 &  $\cap $      &0.13\\
NGC4649 &  20.7 &  41.34  &  10.95 &  -0.35&   $\cap $     &0.15\\
NGC4660 &  20.7 & $<$39.92 & 10.00 &  2.70  &   $\setminus $& \\
NGC4697 &  22.5  & 40.54  &  10.87 &   1.30&  $\setminus $ &0.74\\
NGC4742 &  23.0 & $<$40.21 & 10.14 &  0.41&   $\setminus $ &1.09\\
NGC4874 & 149.3 &  42.28 &  11.50  &  -0.30 & $\cap $      &0.13\\
NGC4889 & 149.3 &  42.20 &  11.60  &  0.01&  $\cap $       &0.05\\
NGC5198 &  51.3 & $<$40.60 &10.53 &      & $\setminus $    &0.88\\
NGC5982 &  59.3 &  41.47 & 10.92  &  -0.80&  $\cap $       &0.11\\
NGC7332 &  32.5 &  $<$40.57 &  10.44 & &  $\setminus $       &0.90\\
NGC7457 &  22.2 & $<$39.99 & 10.18 &  0.00&  $\setminus $    &\\
\noalign{\smallskip}
\hline
\end{tabular} 
\end{flushleft}
\end{table*}
\begin{table*}
\centerline{ Table 1. Continues}
\bigskip
\begin{flushleft}
\begin{tabular}{ l  r  r  r  r  r l }
\noalign{\smallskip}
\hline
\noalign{\smallskip}
 Name   & d$^a$  & log$L_{\rm X}^b$ & log$L_{\rm B}^c$ &
 100$\cdot a_4/a^d$ & Profile$^e$ & $\gamma ^f$\\
   & (Mpc)  & (erg s$^{-1})$ & $(L_{\odot})$ & & & \\
\noalign{\smallskip}
\hline
\noalign{\smallskip}
NGC7626 &  74.3 &  41.58 & 11.11 &  0.20 & $\cap $        &0.21\\
NGC7768 & 164.5 &  42.18 & 11.33  &  0.00 & $\cap $       & 0.00\\
IC1459  &  33.1  & 41.13  &  10.90 &  -0.10&   $\cap $     &0.18\\
\noalign{\smallskip}
\hline
\end{tabular} 
\end{flushleft}
\bigskip
$^a$ Distance, with $H_0=50$ km s$^{-1}$ Mpc$^{-1}$ (see Section 2.1).

$^b$ X-ray luminosity (see Section 2.1).

$^c$ B-band luminosity calculated using the $B_T^0$ values of de
Vaucouleurs et al. (1991) and the distances in the second column.

$^d$ Isophotal shape parameter; $a_4/a$ is the coefficient of the
cosine distortion term expressed as a percentage of major axis 
length. Sources for $a_4/a$ are F97, Bender et al.
(1989), Goudfrooij et al. (1994), Peletier et al. (1990), Franx et
al. (1989), M{\o}ller et al. (1995).  

$^e$ Profile class: $\cap$=core, $\setminus$=power law.

$^f$ Slope of the surface brightness profile in the central regions
($I\propto R^{-\gamma}$). The references for the $\gamma $ values 
are given at the beginning of Section 2.

\end{table*}

\subsection{The X-ray luminosities}

Two corrections need to be applied to the X-ray data of the
literature. The first is a correction for the different passbands:
$Einstein $ based luminosities refer to the 0.2--4 keV energy band,
while the $ROSAT$ based data refer to 0.5--2 keV for the Irwin \&
Sarazin list, to 0.64--2.36 keV for the all-sky survey list, and to
0.4--2.4 keV for Dow \& White (1995). I have transformed all X-ray
luminosities to the (0.2--2) keV band values.  This correction has been
applied by taking into consideration the different spectral parameters
used to derive the X-ray luminosities: thermal emission had been used
with temperatures from 0.7 to 1 keV, Galactic neutral hydrogen
absorption, and metallicities from 0.5 solar to solar.  The amount of
 correction on log \lx ranges from --0.030 for the $Einstein $
luminosities, to +0.072 and +0.071 for the luminosities in the Irwin
\& Sarazin and all-sky survey lists, and to +0.025 for the two Coma
galaxies.

The second correction is the rescaling of the luminosities to the same
distance scale. To maximize the number of galaxies with the same
source for the distance, I have adopted the distances given by Beuing
et al. (1999). For eight galaxies not present in their catalog I have
adopted the distances given by F97, after rescaling (they refer to
$H_0=80$ km s$^{-1}$ Mpc$^{-1}$, while the distances of Beuing et
al. refer to $H_0=50$ km s$^{-1}$ Mpc$^{-1}$).  Table 1 summarizes the
observational information used in this paper.

\section{The relation between \lx and the central brightness profile}

Figure 1 plots the $L_X-L_B$ relation for the galaxies in this sample,
with different symbols for  core and power law galaxies (full and
open circles respectively).  
We note that:\par 1) The least
optically luminous galaxies and the most optically luminous ones are
respectively power law and  core galaxies, as in the sample
studied by F97.
\par 2) The least X-ray luminous
galaxies and the most X-ray luminous ones are again respectively power
law and  core galaxies, consistent with the known
\lx--\lb correlation. For
 the whole {\it Einstein} sample of 148 early
type galaxies  \lx$\propto$\lb$^{1.8-2}$, although with a
large scatter of more than two orders of magnitude in \lx at any fixed
\lb$>3\times 10^{10}L_{\odot}$ (Eskridge et al. 1995). 
For a magnitude-limited sample of early type galaxies with X-ray
emission measured by the $ROSAT$ all-sky survey, Beuing et al. (1999)
find \lx$\propto$\lb$^{2.2 \pm 0.1}$ and a scatter at least as
large as that found by Eskridge et al. (1995).
\par 
3) At intermediate \lb, where they coexist, core galaxies
 span the whole
range of \lx values (roughly two orders of magnitude in \lx), while
power law ones are confined below log \lx (erg s$^{-1})=41$
(hereafter \lup).

It looks as if power law galaxies cannot be more X-ray luminous than
\lup, while  core galaxies show \lx values extending from the
lowest to the highest \lx observed.  In Fig. 2 this result is shown
from the point of view of the relation between \lx and the central
surface brightness slope $\gamma $ (defined in Section 2). A sharp
transition in \lx as $\gamma$ drops below 0.3 (as for  core
galaxies) is clearly seen: the X-ray brightest galaxies are
exclusively  core galaxies, and power law galaxies are never X-ray
brighter than \lup, independently of the $\gamma $ value.

\begin{figure*}
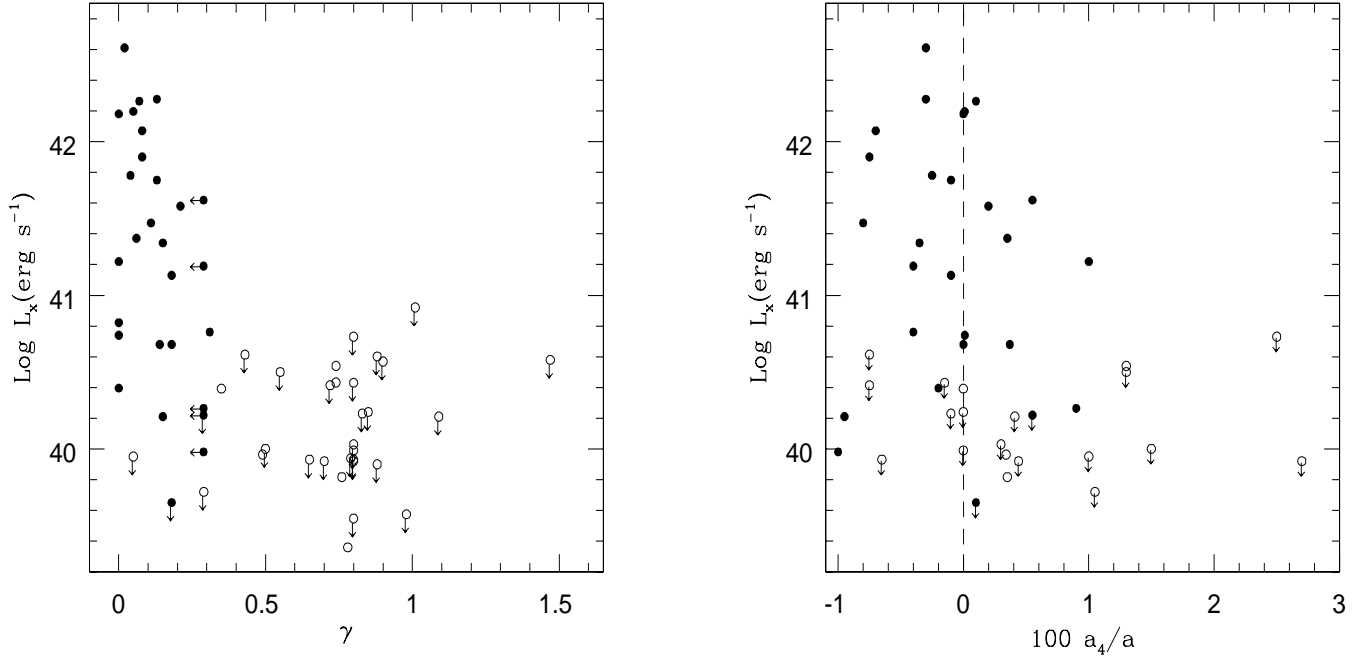

\vskip -4truecm
\hskip -1truecm
\parbox{10cm}{
\psfig{file=9077.f2a,width=14cm,height=15.5cm,angle=0} }
\ \hspace{-4.5cm} \
\parbox{10cm}{
\psfig{file=9077.f2b,width=14cm,height=15.5cm,angle=0} }
\caption[]{The relation between \lx and the slope of the central
surface brightness profile $\gamma$ (left) or the $a_4/a $ parameter
(right) for the galaxies in Table 1. Power law galaxies are shown with
open circles and  core galaxies with full circles. Upper limits
on X-ray luminosities are shown with a downward arrow.  Those galaxies
for which just the classification into core or power law is
given are put respectively at $\gamma =0.3$, with an arrow pointing
leftwards, and at the average $\gamma $ value for their family (i.e.,
0.8, see Section 1).  The vertical dashed line in the \lx--$a_4/a$ plot 
separates boxy and disky galaxies.}
\end{figure*}

\subsection{Statistical analysis}

How strong is the result presented above, from a statistical point of
view? Is the confinement of power law galaxies to \lx$<$\lup a result
of the \lx--\lb correlation, or is it statistically significant in
general, for all of them?  To establish this, I made some statistical
tests for the galaxies in the range of \lb values where power law and
core galaxies coexist.  This range has been chosen close to that
indicated by F97, who find that core and power law profiles coexist
for $-22<M_V<-20.5$ (with $H_0=80$ km s$^{-1}$ Mpc$^{-1}$); by
assuming a B--V=1 and rescaling to $H_0=50$ km s$^{-1}$ Mpc$^{-1}$,
this range corresponds to $10.36<$ log \lb ($L_{\odot})<10.96$.
Considering also the distribution of power law galaxies in Fig. 1, as
overlap range for the present sample I have adopted $10.40\leq $ log
\lb ($L_{\odot}) \leq 10.90$.  In this range there are 15 core
galaxies and 10 power law galaxies.

First I have checked if these two sets of galaxies are consistent with
being drawn from the same distribution function of \lb.  To check this
I have used a variety of two-sample tests [discussed in Feigelson \&
Nelson (1985), and contained in the ASURV package] to verify the null
hypothesis that the two sets are drawn from the same parent
distribution. All these tests give a probability $P\approx 0.4$, from
which one usually concludes that the two sets are consistent with
coming from the same distribution\footnote{A value of $P=0.4$ means
that the observed difference in the \lb distributions can be obtained
in $\sim 40$\% of the cases, using two samples of the same sizes as
those used here, and drawn from the same parent population.}. I obtain
an even higher probability when using the Kolmogorov-Smirnov test
($P=0.57$).

Next, I have repeated the two-sample tests for the \lx values, in
order to check whether power law and core galaxies (again for
$10.40\leq $ log \lb ($L_{\odot}) \leq 10.90$, where they coexist) are
consistent with being drawn also from the same \lx distribution. I
have obtained that the null hypothesis cannot be supported. The
probabilities given by all the tests are around 0.003.  So, the two
sets come from different \lx distributions at the $\sim 3\sigma $
level.

In conclusion, in the \lb range where they coexist, core and power law
 galaxies are similar from the point of view of their optical
 luminosity, while they clearly differ in their X-ray properties.

A linear regression analysis for data sets with censoring in one
variable, with the expectation and maximization (EM) algorithm, and
the Buckley-James algorithm (Isobe et al. 1986), gives a best fit
slope for the \lx--\lb relation of $1.8\pm 0.3$ for core galaxies.
This slope is close to the best fit slope obtained from the analysis
of larger samples (Eskridge et al. 1995, Beuing et al. 1999; see
Section 3).  An estimate of the best fit slope for power law galaxies
is not realistic because of the low number of detections (6 only);
however, in this case a regression analysis gives a best fit slope
consistent with unity.  So, \lx$\propto$\lb could be a good
description of the \lx--\lb relation for power law galaxies, while the
observed deviation of the \lx--\lb relation from a linear one could be
produced by core galaxies.

\begin{figure}
\vskip -5truecm
\hskip -1truecm
\parbox{1cm}{
\psfig{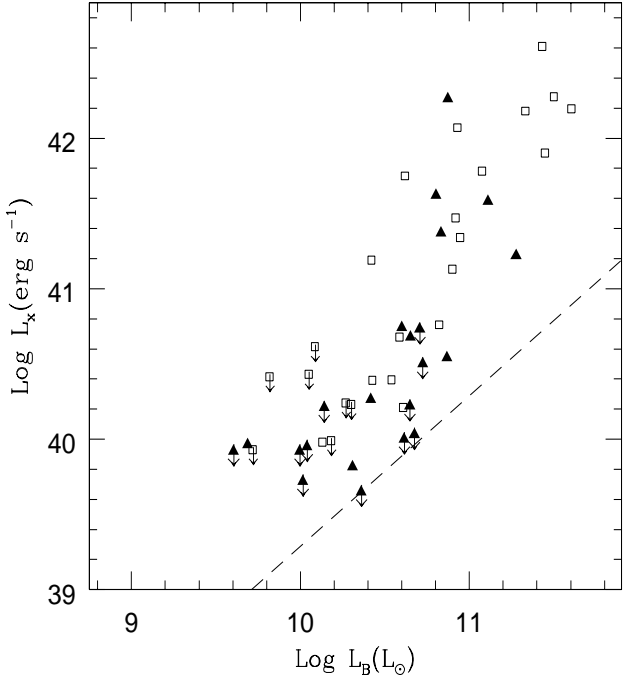} }
\vskip -1truecm
\caption[]{The \lx--\lb diagram for the early type galaxies in Fig. 1,
with isophotal shape measurement available in the literature. Disky
galaxies (those with $100\cdot a_4/a>0.01$) are shown with full triangles
and boxy galaxies with boxes.  The data are taken from Table 1, and the
dashed line is as in Fig. 1.}
\end{figure}

\section{Relation of \lx  with other basic galaxy properties}

Not surprisingly, the behavior of the X-ray emission with respect to
the shape of the central brightness profile is reminiscent of the
trend of \lx with respect to the other basic properties characterizing
early type galaxies, already mentioned in Section 1: one is the
deviation of the isophotal shape from a pure elliptical shape, and is
described by the $a_4/a$ parameter introduced by Bender et al. (1989);
the other is the degree of anisotropy in the velocity dispersion
tensor, as measured by the anisotropy parameter $(v/\sigma)^*$ ($v$ is
the maximum rotational velocity and $\sigma $ is the stellar velocity
dispersion in the central regions; e.g., Davies et al. 1983).  Disky
($a_4>0$) objects are generally low X-ray emitters and flattened by
rotation; boxy ($a_4<0$) and irregular objects show the whole range of
\lx and various degrees of velocity anisotropy (Bender et al. 1989;
see also Fig. 3). The relationship between X-ray emission and $a_4/a$
has been reanalyzed by Eskridge et al. (1995) for the sample of
$Einstein$ galaxies: they confirm that the most X-ray luminous
galaxies are boxy, and find that the trend of \lx with $a_4/a$ is a
wedge, with no highly disky objects at high X-ray emission.
Pellegrini et al. (1997) find similar results for the relation between
\lx and $v/\sigma $ (a measure of the importance of rotation), or
$(v/\sigma)^*$: at high $v/\sigma $ or $(v/\sigma)^*$ there are no
galaxies with high \lx (see also Fig. 4).

What is surprising is that the \lx--$\gamma $ relation is {\it
sharper} than that between \lx and isophotal distortion, or $v/\sigma$
(these are more wedges than L-shapes, as described above). As an
example, I consider here in detail the \lx--$\gamma $ and the
\lx--$a_4/a$ relations (Fig. 2; $a_4/a$ has been measured for 47 of
the galaxies in Table 1).  While the \lx--$\gamma$ relation shows
clearly a dicothomy in the X-ray properties, the \lx--$a_4/a$ relation
looks like a wedge, because not all disky galaxies are low X-ray
emitters.  There are in fact both power law and core galaxies (i.e.,
X-ray faint and X-ray bright objects) for $0\leq 100\cdot a_4/a\leq
1$.  This fact is explained by an inspection of Fig. 3, that is
another version of Fig. 1 where disky and boxy galaxies are plotted
with different symbols: disky and boxy objects are more mixed in
the \lx--\lb relation than core and power law galaxies.  So, Figs. 2
and 3 show that {\it all} power law galaxies have log \lx$<$\lup,
while the same is not true for galaxies with disky isophotes.  One
concludes that a {\it global} property such as \lx {\it is tightly
connected with a nuclear property}, such as the inner profile shape,
and more so than with other characterizing galaxy properties as the
isophotal shape.  This occurs in spite of the fact that there are more
objects in the \lx--$\gamma$ plot than in the \lx--$a_4/a$ one; an
effect of ``blurring'', because a larger sample is considered, is
expected to affect the \lx--$\gamma$ plot more than the \lx--$a_4/a$
one.

It must be noted that the absence of a sharp transition in the
\lx--$a_4/a$ relation might be produced, at least in part, by the
uncertainties associated with the definition of $a_4$.  In fact, the
$a_4$ values given in Table 1 refer to an average between 10 and 60
arcsec, i.e., between typically 0.2 and 1 optical effective radius,
where the shape of the isophotes is known to vary within the
same galaxy, sometimes also significantly (e.g., van den Bosch et
al. 1994).

\section{Discussion}

 The results obtained here indicate that the two families of early
type galaxies recently singled out have also a distinct X-ray behavior
in the (0.2--2) keV band.  The trend between \lx and the shape of the
central profile is similar to the trends between \lx and the isophotal
shape distortion, or the degree of rotational support. However, it is
much sharper than the other two, and it could be at the origin of the
difference in the X-ray properties.

What are the possible causes of this segregation of power
law galaxies to \lx$<$\lup, and how do  core ones evade it?  Can the
change in the shape of the density profile in the central regions be
responsible for such a disparate behavior of \lx?  Or are some other
galaxy properties, to which the change in the shape of the surface
brightness profile is linked, responsible for this difference?  For
example, besides isophotal shape and rotation, distinct evolutions, 
where an important role is played also by massive black holes, could
be responsible for the settling of the two types of profiles (Section 1)
 and also have an effect on \lx.

In the following I speculate on a few conceivable causes of the different
\lx behavior in the two families of early type galaxies. I will focus
first on the direct effect of the inner profile shape  on \lx; then
on that of a massive black hole (MBH); finally on those expected from
possible differences in the environment.  I assume that the
 scatter in \lx is related to  a largely varying quantity of hot
gas within galaxies, as demonstrated observationally by considering
their spectral properties (e.g., Matsumoto et al. 1997).

\subsection{Direct effects of the basic properties of power law
and  core galaxies on their hot gas flows}

Ciotti et al. (1991) simulated the evolution of the hot gas behavior
over the galaxy lifetime for spherical mass distributions having King
profiles (i.e., with flat cores), including type Ia supernova heating.
They showed that the large \lx range observed at fixed \lb can be
produced by small differences in the gravitational potential energy of
the gas, which cause the gas flows to be in the wind, outflow or
inflow stages. Later Pellegrini \& Ciotti (1998) produced a set of
hydrodynamical simulations where the spatial luminosity distribution
at small radii has a power law form ($\propto r^{-2}$ as in the Jaffe
law; this corresponds to the central luminosity density profile of
power law galaxies, Gebhardt et al. 1996).  The resulting gas flows
are mostly partial winds: an inflow develops in the central region,
while the external parts are still degassing.  A large spread in \lx
at fixed \lb is again produced by varying the model input parameters
in their observed ranges, and corresponds to variations in the size of
the inflow region. In particular, for $10.4<$log \lb
$(L_{\odot})<10.9$, \lx values as high as $\sim 2-3\times 10^{41}$ erg
s$^{-1}$ can be easily reached.  So, in the X-ray luminosity values of
a large sample of galaxies there should be no trace of the precise
shape of the mass profile in the very central regions. This is
supported also by the observations that show how all power law
galaxies have low \lx independently of the $\gamma $ values (Fig. 2),
while core galaxies have a large range of \lx with little variation of
$\gamma $.

Galactic rotation is an important property of power law galaxies
(Section 1), and so one might suspect that the simulations mentioned
above are not realistic for these galaxies. However, as shown by
Ciotti \& Pellegrini (1996) and by D'Ercole \& Ciotti (1998) with
simulations, rotation has minor effects on \lx. If no type Ia
supernova's heating is assumed (so that the flows are always cooling
flows) the effect is larger: a reduction of \lx by a factor of six is
obtained when $v/\sigma\approx 0.3$, and a very massive cold disk
forms (of mass $\approx 3 \times 10^{10}M_{\odot}$; Brighenti \&
Mathews 1996). Even though in this case the effect of rotation goes in
the right direction, the observations cannot be fully explained,
because the needed reduction in \lx is at least of one order of
magnitude. Moreover, for $v/\sigma\leq 0.4$ there are both power law
and X-ray bright core galaxies (see Fig. 4), and so the amount of
rotation cannot be taken as the general explanation.

\begin{figure}
\vskip -5truecm
\hskip -1truecm
\parbox{1cm}{
\psfig{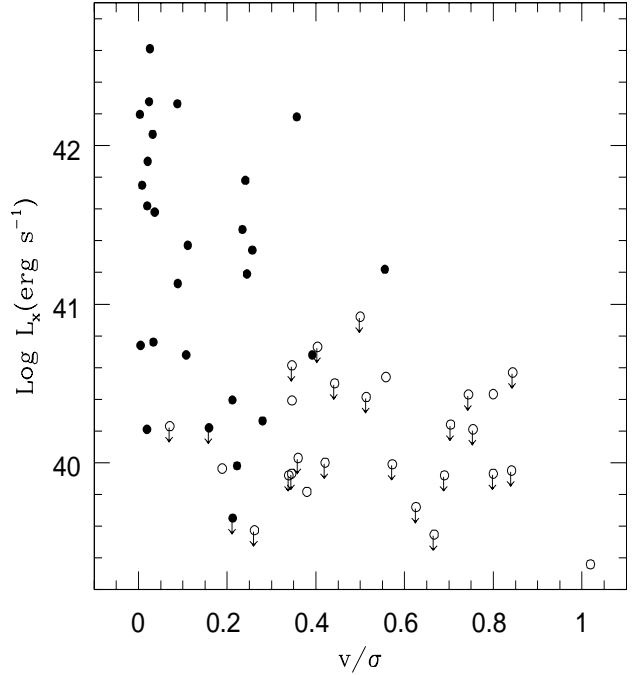} }
\vskip -1truecm
\caption[]{The relation of \lx with the ratio between the maximum
rotational velocity $v$ and the central velocity dispersion $\sigma $
(Section 4).  Power law galaxies are shown with open circles and core
galaxies with full circles. The values of $\sigma $ come from McElroy
et al. (1995); $v$ has been measured for 54 galaxies in Table 1, by
Davies et al. (1983), Franx et al. (1989), Fried \& Illingworth
(1994), Bender et al. (1994), Fisher et al. (1995).}
\end{figure}

A flatter global galaxy shape is another property known to be
associated on average to lower \lx values, from observations (Eskridge
et al. 1995).  The hot gas is less bound when the mass distribution is
flatter, for fixed \lb, so that an outflow is favoured (D'Ercole \&
Ciotti 1998).  This is the case of galaxies where a disk is present,
and might apply preferentially to power law galaxies. Those in the
present sample, for log \lb$(L_{\odot})>10.4$, include four S0s, but
also three roundish (E0-E2) galaxies.  So, again the presence of a
disk has an effect that goes in the right direction, but cannot be
taken as a general explanation (unless all the above mentioned power law
galaxies with rounder shape turned out to be disk galaxies viewed face
on).

In conclusion, large systematic variations of the hot gas content
resulting from differences of the inner profile shape are not
expected, based on numerical simulations; galactic rotation and
flattening of the mass distribution are expected to have an effect,
but seem to be unable to offer a general explanation.  This is
consistent with the \lx--$\gamma$ relation being sharper than the
\lx--$v/\sigma$ one (and the \lx--$\epsilon$ one, where
$\epsilon=1-b/a$ is the projected galaxy ellipticity, Pellegrini et
al. 1997).

\subsection{Effect of the presence of a massive black hole}

When dealing with properties connected with the very central regions
of early type galaxies, one is naturally brought to consider the
effects produced by a central MBH.  In fact, on the basis of various
kinds of observational evidence, in the past few years it has become
commonly accepted that early type galaxies host central massive black
holes, possibly remnants of dead or quiescent QSOs (Kormendy \&
Richstone 1995, Richstone et al. 1998).  Why the X-ray emission in
early type galaxies is not as high as in classical AGNs, due to
accretion of hot gas by the central MBH, is a problem that has been
addressed by Binney \& Tabor (1995), Fabian \& Rees (1995) and by
Ciotti \& Ostriker (1997). Fabian \& Rees suggest that accretion of
gas from the surrounding cooling flow has a very low radiative
efficiency, and that early type galaxies should typically host low
or very low luminosity AGNs.  Ciotti \& Ostriker suggest that the hot
gas can be expelled from the galaxies when a central MBH is
present, because the gas flows are found to be unstable due to Compton
heating. So, most of the time the galaxies are in a low or medium \lx
state, they are in a high \lx state only during global inflows, and
the large observed scatter in \lx is reproduced.

 In both frameworks it is not clear why power law galaxies are found
{\it only} in the low or medium state, while core galaxies span the
whole observed scatter in \lx. A few solutions, where other
ingredients besides the presence of a MBH enter, can be imagined: the
shape of the galaxies (triaxial versus axisymmetric) can have an
effect on the way accretion proceeds. Or the very MBH properties (such
as its mass) could play a role.  In any case if a MBH is important in
determining \lx, the conventional approach to the problem of
interpreting the X-ray properties of early type galaxies requires
revision; the input ingredients are currently just stellar mass loss,
supernova heating, and gas potential energy (see, e.g., Sarazin \&
White 1988, David et al. 1990, Ciotti et al. 1991). More precisely,
since a different amount of hot gas is at the origin of the observed
large scatter in \lx at fixed \lb (e.g., Matusmoto et al.  1997), one
should focus on the relation between central MBH presence and amount
of hot gas.

\subsection{Effect of the environment}

F97 present a preliminary evidence that core galaxies tend to be found
in dense groups and clusters, while power law ones are preferentially
found in the field.  This is in line with the fact that core galaxies
are on average also boxy/irregular (which has been argued to be a
merger signature, or a sign of past accretion and interaction events;
Nieto 1989, Barnes 1992) and show more frequently indications of
accretion of small satellites such as cores within cores, central
counter-rotation, etc.  (Lauer et al. 1995).  In this scenario, what
could be the effects on \lx?  More generally, what is the effect of
the environment on the hot gas content?

An environment denser of intragroup/intracluster gas and of galaxies
should act in the sense of producing both extremely X-ray luminous
galaxies and galaxies with medium or low \lx. In fact accretion of
external gas is needed to explain the X-ray brightest galaxies of the
\lx--\lb diagram (Renzini et al. 1993, Mathews \& Brighenti 1998)
while in the X-ray faint ones the hot gaseous halos could have been
stripped by ambient gas, if it is sufficiently dense, or in encounters
with other galaxies (White \& Sarazin 1991).  This picture would be
consistent with the finding that the larger spread in \lx is shown by
core galaxies, that are preferentially found in high density
environments (and this is so even for the sample studied here).  Also
it has been suggested that MBH presence and galaxy environment might
be connected (F97); in this view cores might result from ejection of stars by
a coalescing black hole binary during a merging event (Quinlan \&
Hernquist 1997, Nakano \& Makino 1999). So, the association of cores
and a large spread in X-ray emission could come from a
combination of environmental effects and MBH presence; for example,
merging produces triaxiality, and this facilitates the gas to reach
the nucleus, and feed the central MBH.  Note that just the variety of
merging and/or tidal interaction conditions (time of the event during
the galaxy history, progenitors' masses and orbits, and so on) might
produce a variety of effects on the hot gas content.  Numerical
simulations are needed to derive more firm conclusions concerning this
aspect; the first results show that the effect of perturbing the hot
gas flows is in the direction of producing a spread in \lx (D'Ercole
et al. 1999).

While various solutions can be imagined to explain the large range of \lx
shown by  core galaxies, again a major problem is why power law
galaxies do not reach high X-ray luminosities, even when they are in a
range of \lb values where  core galaxies can be very X-ray
luminous.  Certainly, they do not seem to experience accretion from
outside, because they do not even show \lx values as high as the
numerical simulations for the hot gas evolution predict for them when
they are isolated (Section 5.1).  So, they should reside in the
field\footnote{The power law galaxies with log \lb$(L_{\odot})\geq
10.4$ of the sample investigated here reside either in the field, or
in groups with 3 to 18 members, or in Virgo (NGC4621).}, or never
be the central dominant galaxies in clusters, subclusters and
groups. But this seems not to be the case: for example NGC1553,
NGC4697 and NGC5198 are also the brightest members of their groups.
So, the problem remains of why power law galaxies do not become very
X-ray bright, even when they are optically dominant.

\section{Conclusions}

I have investigated the relationship between X-ray emission and shape
 of the inner surface brightness profile, for a sample 59 early type
 galaxies. I have found that:

$\bullet $ The family of core galaxies spans the whole observed range
 of \lx values (two orders of magnitude in \lx) while that of power
 law galaxies is confined to log \lx (erg s$^{-1})<41$.  This
 dicothomy in the X-ray properties holds even in the \lb range where
 the two families coexist.
 
$\bullet $ The relation between \lx and the shape of the inner profile
is sharp, and is stronger than the relations of \lx with the other
basic properties characterizing the two families of early type
galaxies.  For example, being a power law galaxy
is connected with a low \lx with no exception, while the same cannot
be said for the property of having disky isophotes. So, a global
quantity such as \lx is surprisingly tightly connected with a nuclear
galaxy property.

$\bullet $ A linear regression analysis shows that 
 \lx$\propto$\lb could be a good description of the \lx--\lb relation for
 power law galaxies, while core galaxies deviate from this relation.

$\bullet $ Different possible reasons can be argued  for the origin of
the dicothomy in the \lx behavior.  The central profile shape itself
should not be the main factor, given the results of previous numerical
simulations of the hot gas evolution and the absence of a trend of
\lx with the $\gamma $ values.  A higher degree of rotation of power
law galaxies could produce an effect, but not large enough, and, much
like the possibility of  a higher degree of flattening
of the mass distribution, does not seem to  be a general explanation
of the reduction in \lx for the whole class of power law galaxies.

$\bullet $ It has been suggested that nuclear massive black holes and
environmental differences are important for explaining the dicothomy
of the inner light profiles; both aspects are likely to influence the
hot gas content.  While a few explanations can be imagined for the
large spread in the X-ray luminosities of core galaxies, a clearly
open problem is why power law galaxies never become X-ray bright, even
when they are the brightest objects of the groups where they reside.

$\bullet $ If a massive black hole and the environment have a
fundamental role in determining \lx, the problem of
interpreting the X-ray properties of early type galaxies becomes much
more complex than thought so far, when the input ingredients were just
stellar mass loss, supernova heating and gas potential energy.

\begin{acknowledgements}
I am grateful to G. Bertin, J. Binney, L. Ciotti, L. Ferrarese
and R. Sancisi for useful comments, and to G. Zamorani for advice on
statistical analysis. ASI and MURST (contract CoFin98) are acknowledged 
for financial support.

\end{acknowledgements}

\end{document}